# Machine learning-guided synthesis of advanced inorganic materials

*Bijun Tang[1†], Yuhao Lu[2†], Jiadong Zhou[1], Han Wang[1], Prafful Golani[1], Manzhang Xu[1], Quan Xu[3], Cuntai Guan[2]\*, and Zheng Liu[1,4,5,6]\**

[1]School of Materials Science and Engineering, Nanyang Technological University, Singapore 639798, Singapore.

[2]School of Computer Science and Engineering, Nanyang Technological University, Singapore 639798, Singapore.

[3]State Key Laboratory of Heavy Oil Processing, China University of Petroleum (Beijing), Beijing, China

[4]Centre for Micro-/Nano-electronics (NOVITAS), School of Electrical & Electronic Engineering, Nanyang Technological University, 50 Nanyang Avenue, Singapore 639798, Singapore

[5]CINTRA CNRS/NTU/THALES, UMI 3288, Research Techno Plaza, 50 Nanyang Drive, Border X Block, Level 6, Singapore 637553, Singapore

[6]Chemistry and Materials Centre, Nanyang Environment and Water Research Institute, Singapore 637141, Singapore

\* Email: CTGuan@ntu.edu.sg, z.liu@ntu.edu.sg

**Synthesis of advanced inorganic materials with minimum number of trials is of paramount importance towards the acceleration of inorganic materials development. The enormous complexity involved in existing multi-variable synthesis methods leads to high uncertainty, numerous trials and exorbitant cost. Recently, machine learning (ML) has demonstrated tremendous potential for material research. Here, we report the application of ML to optimize and accelerate material synthesis process in two representative multi-variable systems. A classification ML model on chemical vapor deposition-grown $MoS_2$ is established, capable of optimizing the synthesis conditions to**

**achieve higher success rate. While a regression model is constructed on the hydrothermal-synthesized carbon quantum dots, to enhance the process-related properties such as the photoluminescence quantum yield. Progressive adaptive model is further developed, aiming to involve ML at the beginning stage of new material synthesis. Optimization of the experimental outcome with minimized number of trials can be achieved with the effective feedback loops. This work serves as proof of concept revealing the feasibility and remarkable capability of ML to facilitate the synthesis of inorganic materials, and opens up a new window for accelerating material development.**

Material synthesis is always a challenging problem hindering the development of advanced inorganic materials. Complex synthesis processes not only possess various uncertainties but are also costly and time-consuming[1]. Taking the synthesis of two-dimensional (2D) materials as an example, 2D materials have been spotlighted in recent years attributed to their unique and fascinating properties[2-5], and chemical vapor deposition (CVD) is considered as one of the most promising methods to realize the controllable and scalable synthesis[6-8]. However, CVD process contains numerous variables like reaction temperature, chamber pressure, carrier gas flow rate, etc., significantly aggravating its unpredictability. Especially, early exploration of the optimal synthesis condition was solely driven by a laborious trial-and-error process, rendering extremely long development cycle. Additionally, large number of trials together with expensive precursors and high energy consumption result in exorbitant research and development cost. Not only CVD, other multi-variable synthesis methods including hydrothermal, chemical vapor transport (CVT), atomic layer deposition (ALD) and molecular beam epitaxy (MBE), all have such issues. Therefore, an effective learning strategy towards optimizing and accelerating the synthesis of advanced inorganic materials, is urgently required.

Recently, the emergence of contemporary machine learning (ML) methods has demonstrated the great potential of statistical algorithms to substantially accelerate the materials development, as shown in Fig. 1a. For instance, ML models have been successfully applied for predicting new perovskite halides[9], metallic glasses[10], properties of inorganic materials and grain boundary energies of crystalline materials[11, 12], as well as identifying the material phase transition and crystal structures[13, 14]. However, most of the work reported are limited to phase 2, exploration of new materials and prediction of properties[15-19]; while phase 3, material synthesis, the critical step towards the final application of materials, remains less studied. Moreover, along with development of high-throughput first-principles computations, the need for efficient and controllable synthesis become even more pressing to cope with the dramatically growing volume of predicted and screened materials. Among the few pioneering studies of ML-guided synthesis, most of them focus on exploring the space and underlying mechanism of chemical reactions in the reactors, instead of the large-scale material synthesis[20-22]. Thus, it is timely to explore the capability of ML to guide the synthesis process of advanced materials.

Here, to demonstrate the feasibility of optimizing and accelerating the materials' synthesis process through ML, we implement supervised ML for the CVD synthesis of 2D $MoS_2$, which is promising candidate for numerous applications[2, 3, 23-25]. The paradigm is schematically depicted in Fig. 1b. To optimize the synthesis condition, CVD synthesis data are collected experimentally, and a classification model is then constructed, capable of predicting the probability of successful synthesis given a set of CVD parameters and recommending the most favorable conditions. To accelerate the development of new materials, progressive adaptive model (PAM) is further introduced, which can effectively maximize the experimental success rate and reduce the number of trials. Most importantly, the principle demonstrated on CVD synthesis can be extended to other multi-variable synthesis methods, not only to improve the success rate, but also to enhance the process-related properties. A regression model has been successfully constructed for the hydrothermal-grown carbon quantum dots (CQDs), to enhance the property of photoluminescence quantum yield (PLQY).

**ML-Guided CVD Synthesis with High Success Rate**

To realize the controllable synthesis of advanced inorganic materials, the ambiguous relationship between various synthesis parameters and outcomes need to be understood. ML possesses great potential of unveiling such relationship through studying existing synthesis data, and then recommends optimal growth conditions with high success rate. Here, the CVD-grown $MoS_2$ is targeted not only because of its fascinating 2D properties; more importantly, the information obtained and methodology established with $MoS_2$ can be readily extended to other CVD-grown 2D materials, especially the transition metal dichalcogenides (TMDs), as they are synthesized in a similar manner. Controllable growth is crucial for the next-generation 2D electronics, as it provides the material basis.

**Dataset.** The experimental setup for $MoS_2$ growth is schematically illustrated in Supplementary Fig. 1a. Detailed information on the synthesis process is shown in Methods. 300 experiments were carried out in the laboratory with different combinations of synthesis parameters. Among them, MoS$_2$ were successfully obtained in 183 experiments (61%), whereas the rest 117 experiments showed negative results (39%). A binary classification model was thus developed by defining "Can grow" as positive class and "Cannot grow" as negative class. Characterizations of a typical "Can grow" sample including optical microscopy (OM), Raman spectroscopy and scanning transmission electron microscopy are shown in Supplementary Fig. 1b-e. If no MoS$_2$ flakes can be observed with OM or the size of the MoS$_2$ flake is smaller than 1 μm, which has no practical use, then it is considered as "Cannot grow". In addition, owing to the resolution limit of OM, it is hard to determine whether the point of interest is the sample or a nucleation site when its size is smaller than 1 μm.

**Feature Engineering.** 19 initial features (Supplementary Table 1) including gas flow rate, reaction temperature and reaction time were chosen to describe the CVD process collectively. After eliminating the fixed parameters and those with missing data, 7 features were retained and constituted the final feature set. The new feature set consisted of distance of S outside furnace (D), gas flow rate ($R_f$), ramp time ($t_r$), reaction temperature (T), reaction time (t), boat configuration (F/T) where F and T represent flat and tilted, and addition of NaCl. Detailed feature overview is shown in Supplementary Table 2. Pearson correlation coefficients were calculated to identify the positive and negative correlations between pairs of features (Fig. 2a). Low linear correlations for most of the features indicate that independent and informative features have been selected to form the essential feature set[18].

**Model Selection.** Based on "no-free-lunch theorem"[26], there is no universally optimal algorithm for all problems. Thus, in this work, XGBoost classifier (XGBoost-C, a more powerful variant of gradient boosting decision tree; see Methods)[27], support vector machine classifier (SVM-C)[28], Naïve Bayes classifier (NB-C)[29], and multilayer perceptron classifier (MLP-C)[30] were employed on MoS$_2$ dataset for selecting the best model. Each candidate model was evaluated with ten runs of nested cross validation to avoid overfitting in model selection[31]. Detailed working principle is schematically illustrated in Supplementary Fig. 2. The whole dataset is shuffled in each run of nested cross validation, with the outer loop assessing the performance of the models on unseen data sets (ten-fold outer cross validation), and the inner loop conducting hyperparameter search and model fitting (ten-fold inner cross validation).

Quantitative comparisons among models are shown in Supplementary Fig. 3, indicating that XGBoost-C reproduces the best agreement to the true synthesis outcomes and possesses the most stable testing performance. Receiver operating characteristic (ROC) curve of XGBoost-C is presented in Fig. 2b, which reports the prediction performance for positive class (correctly versus incorrectly predicted) with all possible prediction thresholds (see Methods)[32]. Large area under ROC curve (AUROC) of 0.96 suggests the model's great capability to distinguish between "Can grow" and "Cannot grow" classes. Moreover, the learning curve displayed in Supplementary Fig. 4 shows the reducing gap between training and cross validation performances along with the increase of training set. It denotes that XGBoost-C is not overfitting, contributing to its generalizability. As a result, XGBoost-C was chosen to learn the nonlinear mapping from CVD synthesis parameters to experimental outcome from the whole MoS$_2$ dataset, and subsequently made predictions for unexplored conditions.

**Optimization of synthesis condition for higher success rate.** SHapley Additive exPlanations (SHAP) was used to unveil the intricate relationship between features and output captured in the as-obtained best model, XGBoost-C (see Methods). It is a unified approach to interpret ML models by additive feature importance measures that is proven to be unique and consistent with human intuition[33]. As shown in Fig. 2c, the gas flow rate ($R_f$) plays the most important role in determining whether $MoS_2$ can be synthesized, followed by the reaction temperature (T) and reaction time (t). This is in good agreement with laboratory experience. $R_f$ is a very important growth parameter, which affects the exposure time and S source controlling. At very low $R_f$, $MoS_2$ can barely be grown as few precursors are transported to the growth substrate. $MoS_2$ can hardly be synthesized at very high $R_f$ either, because too many precursors are transferred downstream to the end of the furnace tube instead of the growth substrate. T, on the other hand, is critical in determining the vapor pressure of the reactants and affecting whether activities of nucleation and grain growth can happen. Moreover, the thickness of as-formed sample normally possesses a positive correlation with t. Thus, the appropriate selection of $R_f$, T and t is of great importance for the synthesis of atomic-layer 2D $MoS_2$ both theoretically and experimentally[6, 34, 35]. This can also serve as a general guidance for the synthesis of other 2D materials, especially TMDs (Supplementary Fig. 5).

The optimal synthesis conditions of 2D $MoS_2$ were further identified with XGBoost-C in the unexplored search space. To achieve this, we firstly defined the possible input range of each critical parameter as shown in Table 1, resulting in 52,920 possible combinations in total. Small step sizes were employed for $R_f$, T and t attributed to their significant importance interpreted from the model, as their subtle changes might lead to substantially different output. Next, XGBoost-C was applied to predict the "Can grow" probability of all 52,920 conditions. 10 synthesis conditions with the highest predicted probabilities were then tested in laboratory with results shown in Supplementary Table 3. 2D $MoS_2$ were successfully synthesized under all 10 conditions, which substantially exceeds the 61% success rate in the original $MoS_2$ dataset, verifying the validity and effectiveness of this model in real-world data.

**Acceleration with progressive adaptive model (PAM).** With the proven effectiveness of ML guidance in material synthesis, we hypothesized that the early intervene of ML might lead to an enhanced success rate and time reduction. Therefore, progressive adaptive model (PAM) was further proposed, which started from small initial dataset and evolved with iterative feedback loops. The performance of PAM was investigated with the same CVD-grown $MoS_2$ dataset through off-line analysis.

The schematic of PAM is provided in Fig. 3a. Initially, $N_1$ synthesis conditions were randomly chosen and labeled by their respective synthesis outcomes extracted from the dataset. $N_1$ was determined such that there are at least ten samples in each class to draw the boundary between classes, in order to perform the ten-fold cross validation. XGBoost-C model was first trained on $N_1$ data and then predicted the "Can grow" probability of the rest (300 - $N_1$) synthesis conditions, assuming the experiments were yet to be conducted. The condition with the highest probability together with its true label were then augmented to the training set. The same steps were repeated in the subsequent loops. PAM stopped at the critical point, $N_c$, where the "Can grow" probability of all (300 – $N_C$) conditions were predicted to be smaller than 50.0 % for the first time (i.e. PAM predicted all the remaining conditions as "Cannot grow"). The success rate based on all the experiments conducted was evaluated to assess the performance of PAM-guided materials synthesis. One typical trial of PAM was visualized and analyzed in Supplementary Fig. 6.

To further investigate whether the randomly selected initial training set will affect the model's performance, PAM was repeated 1000 times on shuffled $MoS_2$ dataset. Detailed validation for the selection of 1000 trials is provided in Supplementary Fig. 7. 1000 trials of PAM resulted in a distribution of $N_C$ as shown in Fig. 3b, where $N_C$ mainly clusters to the mean 189.28 ($\pm$28.89). The respective success rate, time reduction and true positive rate of 1000 trials were calculated and presented in Fig. 3c (see Methods). Success rate maintains stably high at around 83.60%, with very little variance ($\pm$5.57). Average of time reduction and true positive rate are 36.90% ($\pm$9.63) and 87.13% ($\pm$13.78) respectively. It is worth mentioning that the trade-off between large time reduction and high true positive rate is inevitable. Overall, PAM can greatly improve efficiency of iterative experiments, achieving considerable time reduction.

Based on the CVD-grown $MoS_2$ dataset, through the model construction, optimization and PAM, we have demonstrated that proposed ML methodology provides a possibility of achieving high success rate and time reduction and has great advantages in navigating complex multi-variable synthesis systems of inorganic materials.

**ML-guided hydrothermal synthesis with enhanced targeted property**

To further test the generalizability of our established methodology in handling various inorganic material synthesis problems, we have applied ML model on the hydrothermal system aiming to enhance the process-related properties as shown in Figure 1b. Hydrothermal is another commonly used multi-variable method to obtain inorganic materials. Recently CQDs obtained by hydrothermal method have gained substantial attention for their tunable low toxicity, high biocompatibility and robust surface engineering capacity and thus have been widely used in diverse fields including sensors, catalysis, bio-imaging, and energy harvesting etc[36, 37]. Therefore, improving the properties of CQDs with ML is of great research interest. Most importantly, it showcases the feasibility of our methodology in addressing regression problems on top of classification problems (CVD-grown $MoS_2$ dataset).

**Dataset and Model Construction.** For the growth of CQDs, the experiment setup and detailed synthesis process are provided in Supplementary Fig. 8 and Methods. Empirically, six hydrothermal parameters were identified as significant input features: pH value (pH), reaction temperature (T), reaction time (t), mass of precursor A (M), ramp rate ($R_r$) and solution volume (V). Detailed feature overview is shown in Supplementary Table 4. Feature correlation is presented in Fig. 4a, with low linear correlations verifying the great effectiveness of feature selection. As high PLQY is a key property for quantum dots desired for applications including bio-images, biosensors, W-LEDs, photocatalysis etc., it was targeted as the output here. 467 experiments were carried out in the laboratory with different combinations of growth parameters, and respective PLQY ranging from 0 to 1 were recorded.

To best infer PLQY from the features, several regression algorithms were evaluated with nested cross validation mentioned above, including XGBoost regressor (XGBoost-R)[27], support vector machine regressor (SVM-R)[28], Gaussian process regressor (GP-R)[38], and multilayer perceptron regressor (MLP-R)[30]. Coefficient of determination ($R^2$) was adopted as the primary performance indicator, which measures the proportion of variance of the outcome (i.e. PLQY) that is predictable from the features. Detailed model comparison results are provided in

Supplementary Fig. 9. XGBoost-R outperforms the rest by a large margin with its $R^2$ equals to 0.8402, where approaching one is desirable; and thus selected as the best model.

**Optimization for higher PLQY.** After obtaining the trained XGBoost-R model with the full dataset, feature importance of the hydrothermal system has been studied as well. As shown in Fig. 4b, pH value plays the most important role in determining the value of PLQY, followed by reaction temperature and reaction time. This coincides with our expectation: 1) pH will affect the formation of CQDs, as small stable CQDs would dissolve in the acidic and basic solutions. 2) Optimal reaction temperature is required for the formation of CQDS. Higher temperature will result in higher average kinetic energy of molecules and more collisions per unit time, damaging the formation of stable CQDS; and lower temperature would slow down or even cannot initiate the formation of CQDS because of insufficient chemical reaction energy. 3) Reaction time is an important factor controlling the size of CQDS, which then affect their photoluminescence properties owing to quantum confinement effect. Inadequate time will not lead to the formation of CQDS, while prolonged time will result in large-sized CQDS. When the size is larger than the exciton Bohr radius, the quantum confinement effect of CQDS will be impaired and PLQY will be reduced.

The trained XGBoost-R model was then applied to predict the PLQY of 1,555,840 possible synthesis conditions resulted from the combinations of different values of features shown in Table 2. 11 unexplored synthesis conditions were recommended by the model attributed to their highest predicted PLQY. Experiments were then carried out in the lab, and high PLQY of 55.5% (vs. 52.8%, the highest PLQY in the training set) was achieved surprisingly, which is one of the highest PLQY reported with such ultra-low heteroatom doping precursor ratio[36]. Characterizations of the as-obtained CQDs are provided in Supplementary Fig. 10. Moreover, the average PLQY in the recommendation set reaches 53.56, more than twice of the average value of the training set. Comparison of the performances of the training set and the ML-provided recommendation set is shown in Supplementary Fig. 11, indicating great effectiveness of the ML model.

**Acceleration with PAM.** The limitation with traditional experimental exploration arises from randomness due to the lack of guidance. Specifically, the optimal synthesis condition within the pre-defined search space need to be explored through large number of experiments. In the CQDs dataset, without ML guidance, the probability of finding the best synthesis condition is evenly distributed among the full dataset of 467 experiments, leading to excessive waste of time. To tackle such problems as well as to test the generalizability of PAM proposed above, the performance of PAM on CQDs regression dataset was carefully examined, aiming to efficiently identify the best synthesis condition with minimized number of trials.

In a typical run of PAM, along with the increase of explored conditions, the corresponding true yield shows a clear declining trend, suggesting that PAM model is capable to identify the best synthesis conditions at the early stage of PAM loops (see Fig. 4c). In contrast, the original experimental exploration possesses a much more random nature. In order to verify that PAM can perform stably with varying initial training sets, PAM on CQDs dataset was repeated for 1000 times with randomly chosen initial training sets. In each trial, the loop number when the best synthesis condition is found, denoted by $N_c$, was recorded. The results of 1000 trials were summarized in Fig. 4d, from which we can see that the PAM model is able to find the best condition of this confined dataset within 115 experiments with 99.9% confidence. Comparing with the 467 experiments through trial-and-error, 75.37% time reduction has been achieved with the help of PAM.

**Conclusion and outlook**

In summary, ML has been successfully applied to guide the synthesis of inorganic materials. High AUROC of 0.96 was achieved with XGBoost-C for the CVD system, to predict the synthesis result of 2D $MoS_2$ and optimize its CVD synthesis condition. PAM has further been built, whose active feedback loop renders ML capable to guide new material synthesis at the beginning stage, to enhance the experimental outcome as well as minimize the number of trials. More importantly, the proposed methodology could be extended to any type of multi-variable synthesis methods across different material categories, which have been again applied in the hydrothermal system, to effectively improve the process-related properties (i.e. PLQY) of CQDs. Our results demonstrate the great capability and potential of ML to optimize and accelerate the material synthesis process, promoting the development of advanced inorganic materials for practical applications in terms of time reduction and property enhancement.

The primary target of this work is to test the feasibility of introducing ML to guide material synthesis. Satisfying results achieved in both CVD and hydrothermal synthesis systems have unveiled great potential and effectiveness of the proposed. ML models consisting of one type of material with a few features are adopted at the current stage to simplify the complex problem through bypassing the chemistry factors behind the material synthesis process. However, to further exploit the useful information contained in historical trials and effectively guide material synthesis, a more comprehensive model involving chemistry-related features such as the vapor pressure, solubility, reactivity etc., as well as various types of material is very much required. It will not only give more accurate predictions and guidance, but also reveal new information or hypothesis regarding the fundamental mechanism of successful synthesis through inverting the model.

**Methods**

**Synthesis of MoS$_2$.** Sulfur (S) and molybdenum trioxide (MoO$_3$) were used as precursors. Si wafer with a 280 nm SiO$_2$ top layer was used as substrate. The MoO$_3$ powder was put into the boat, and Si/SiO$_2$ substrate was put on the boat with the polished surface down. The boat was then placed in the middle of the 1-in. diameter quartz tube. Sulfur powder was positioned a few centimeters away from the furnace mouth in the upstream and Argon (Ar) gas was used as the carrier gas. The system was heated to the growth temperature with designated ramping rate and maintained for a few minutes for the growth of MoS$_2$.

**Synthesis of carbon quantum dots (CQDs).** 10-60 mL 0.01 M sulfamide solution and 0.2-20 g sodium citrate were added into a 100 mL Teflon-lined stainless-steel autoclave. Then, the autoclave was kept in an oven at 80-300 °C for 0.1-12 h. After the reaction, the resulting product was filtered using a 0.22 mm membrane filter followed by concentrating using rotary evaporator to obtain the purified (S, N)-CQDs. The filtrate was dialyzed in a 500 Da dialysis bag for 2 days to obtain the final S,N-CQDs, against ultrapure water which was renewed every

10–12 h, until almost no Na+ (below detective limit) was detected in DI water.

**XGBoost.** XGBoost derived from Gradient Boosting Decision Tree (GBDT) [39, 40], is a typical class of gradient boosting that employs decision trees as base estimators. It makes decision through an ensemble of M base estimators $h_m, m = 1, \dots, M$:

$$\hat{y}_i = \sum_{m=1}^{M} h_m(x_i)$$

Given N training data $\{(x_i, y_i)\}_{i=1}^{N}$, the objective is to minimize:

$$obj(\theta) = \sum_{i=1}^{N} l(y_i, \hat{y}_i) + \sum_{m=1}^{M} \Omega(h_m)$$

where $\sum_{i=1}^{N} l(y_i, \hat{y}_i)$ is the training loss and $\sum_{m=1}^{M} \Omega(h_m)$ is a regularization term which penalizes complexity of the base estimators. Additive training strategy adds one new tree at a time, by choosing the tree that optimizes the objective at step t: $obj(\theta)^t = \sum_{i=1}^{N} l(y_i, \hat{y}_i^t) + \sum_{m=1}^{t} \Omega(h_m)$ whereas $\hat{y}_i^t = \sum_{m=1}^{t} h_m(x_i) = \hat{y}_i^{t-1} + h_t(x_i)$.

**ROC Curve.** To plot the ROC curve for XGBoost-C model[32], nested cross validation is employed to generate predicted probabilities for 300 data samples respectively. In ten-fold outer cross validation, nine folds are used as model development set, while the predicted probabilities of the remaining 30 samples are recorded. In inner cross validation, the best hyperparameters are determined on model development set with stratified ten-fold cross validation. Sensitivity, y-axis of ROC curve, indicates the percentage of true positive samples that are correctly predicted. Specificity is the percentage of true negatives that are correctly predicted[41]. 1−specificity, x-axis of ROC curve, is the percentage of true negative samples that are falsely predicted as positive.

**SHapley Additive exPlanations (SHAP).** SHAP is a unified approach for additive feature attribution, which produces theoretically sound and unique solutions[33]. The explanation model $g(z')$ satisfies $g(z') = \phi_0 + \sum_{i=1}^{M} \phi_i z'_i$, whereas $z' \in \{0,1\}^M$, M is the number of input features, and $\phi_i \in \mathbb{R}$. $z'_i = 1$ indicates a feature that is being observed, otherwise it is denoted by 0. $\phi_i$ represents the feature importance value.

To compute SHAP values, $f_x(S) = E[f(x)|x_S]$ is defined where $f$ is the function or ML model to be explained, $S$ is the set of non-zero indexes in $z'$, and $E[f(x)|x_S]$ is the expected value of the function conditioned on the subset $S$ of the input features. Using these conditional expectations, SHAP value is assigned to each feature:

$$\phi_i = \sum_{S \subseteq N \setminus \{i\}} \frac{|S|!(M-|S|-1)!}{M!} [f_x(S \cup \{i\}) - f_x(S)]$$

where $N$ is the set of all input features.

**True positive rate, success rate and time reduction on the whole dataset.** True positive rate is defined as the number of correctly predicted positive over the total number of true positive samples. The true labels are from experimental results, with positive and negative class referring to "Can grow" and "Cannot grow" respectively. At $N_C$ of each trial of PAM, the true positive is computed as the number of "Can grow" conditions explored divided by the total number of true conditions in the whole dataset (i.e. 183 for the $MoS_2$ dataset). Success rate is defined as the number of "Can grow" conditions explored divided by the total number of explored conditions (i.e. $N_C$), while time reduction is calculated as $\frac{|N_{all} - N_c|}{|N_{all}|}$.

**Acknowledgments**

This work was supported by the Singapore National Research Foundation under NRF award number NRF-NRFF2013-08, Tier 2 MOE2016-T2-2-153, MOE2016-T2-1-131, MOE2015-T2-2-007, Tier 1 RG4/17.

**Author Contributions**

B. T and Y. L contributed equally to this work. Z. L and C. G conceived and supervised the project. B. T and Y. L developed and implemented the method, as well as wrote the paper. J. Z and M. X performed the chemical vapor deposition experiments. P. G collected the data and H. W constructed preliminary ML models. Q. X conducted hydrothermal experiments. All authors discussed the results and commented on the manuscript.

**Competing Financial Interests**

The authors declare no competing financial interests.

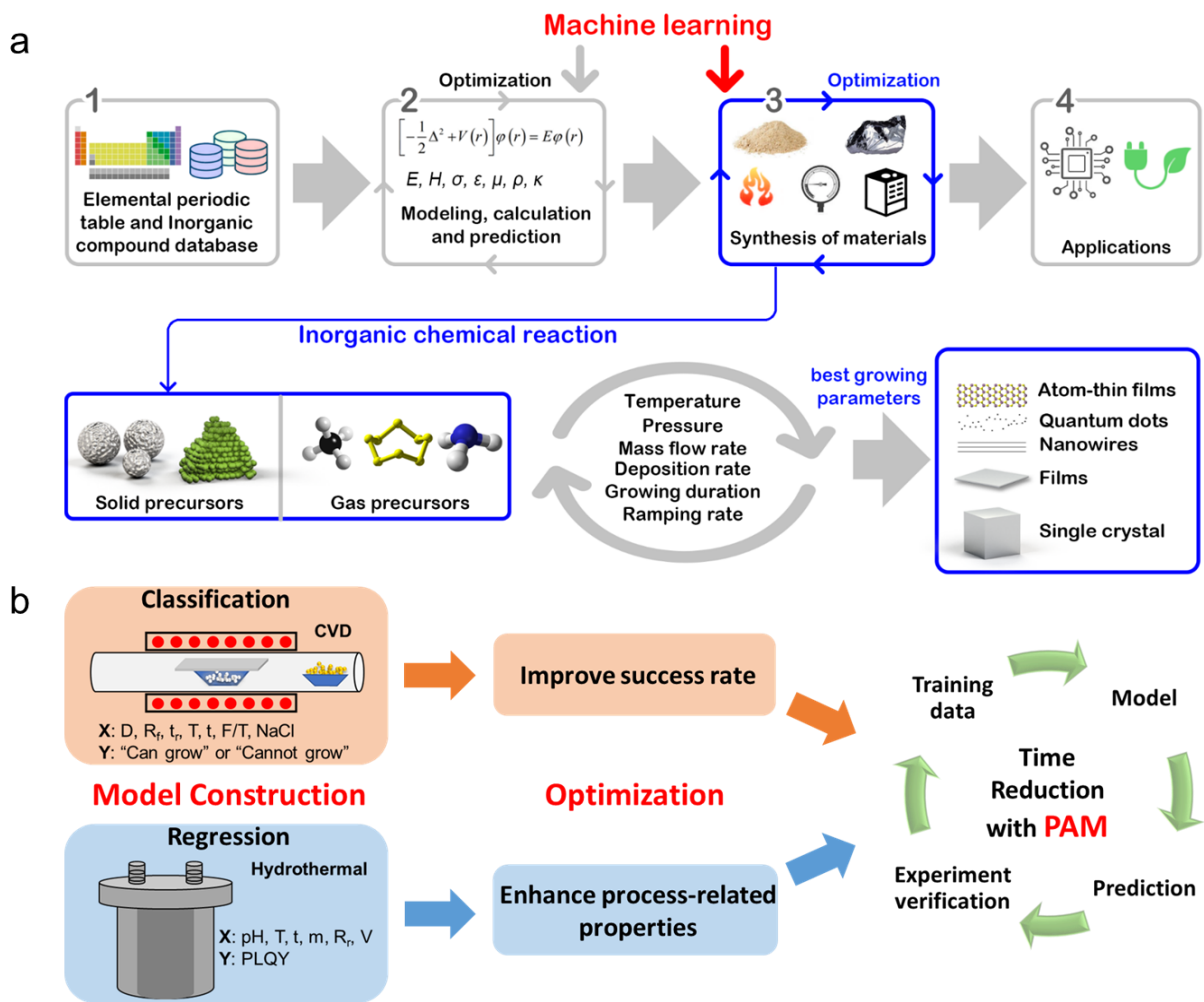

**Figure 1 | Schematic illustration of a paradigm for optimized and accelerated inorganic material synthesis through iterative combination of ML with experimentation. a**, The life cycle of materials development includes four phases: elements and compound database preparation, property prediction and optimization, materials synthesis, as well as practical application. As ML methods has demonstrated its great potential in the second phase, its feasibility in material synthesis scenario is further investigated in this work. **b,** Workflow to achieve the optimization and acceleration of inorganic material synthesis. Model construction, optimization and progressive adaptive model (PAM) are the three key steps, applicable to both classification and regression material synthesis scenarios.

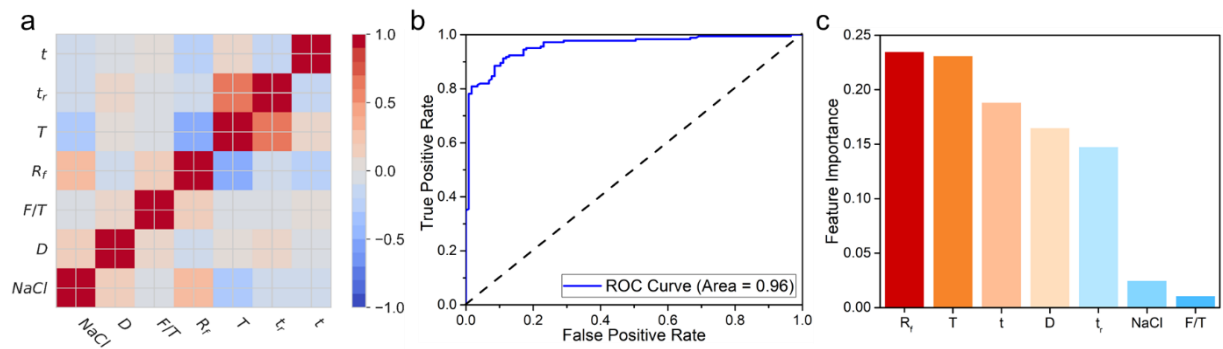

**Figure 2 | Model evaluation and interpretation for optimizing CVD-grown MoS₂ synthesis conditions. a**, The heat map of the Pearson correlation coefficient matrix among the selected features for CVD-grown MoS₂. **b**, Receiver operating characteristic (ROC) curve for XGBoost-C. High AUROC unveils the great capability of the model to distinguish between two classes. **c**, Feature importance retrieved from XGBoost-C that learns from all 300 data samples, computed through unique and consistent SHapley Additive exPlanations (SHAP) method. $R_f$ and T are the two most important features.

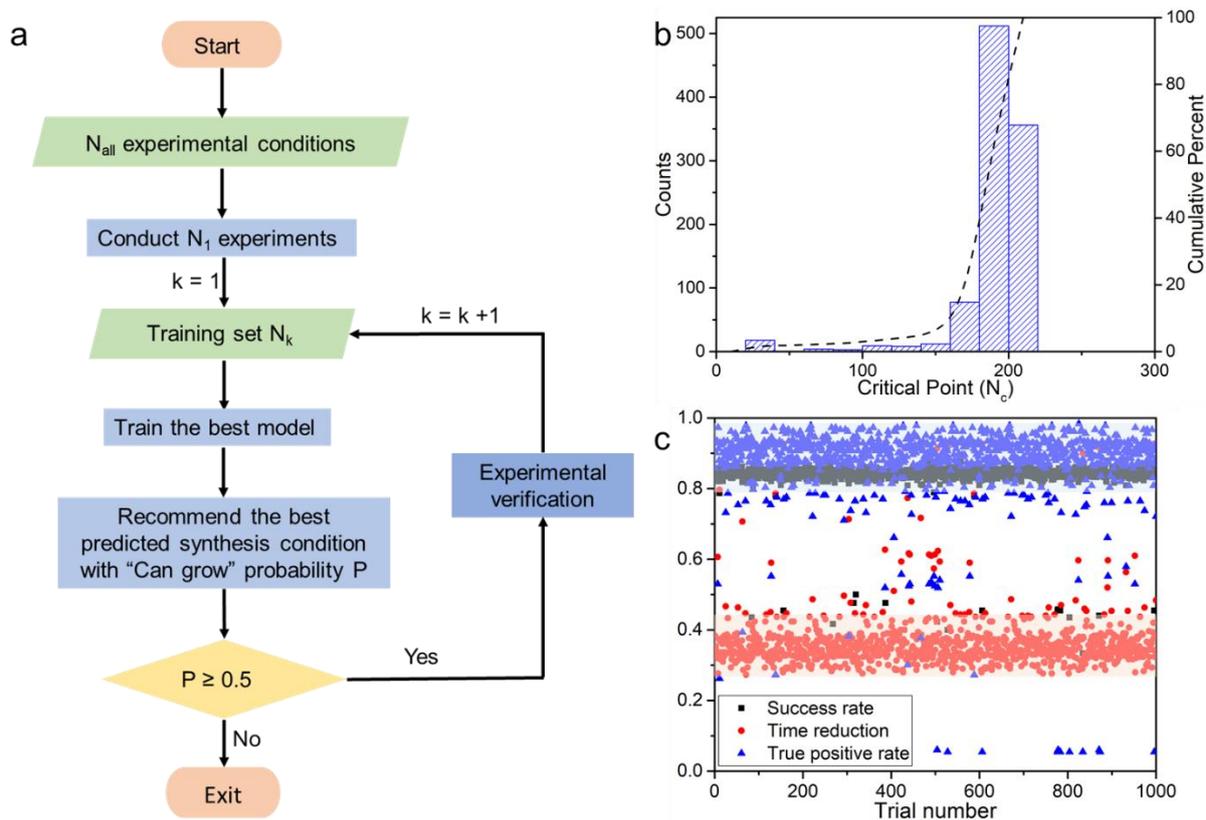

**Figure 3 | Schematic of progressive adaptive model (PAM) for accelerating inorganic synthesis and overall performance on CVD-grown MoS$_2$. a**, Outline of the PAM work-flow, which displays feedback loops and exiting condition. **b**, Distribution of the critical points of 1000 PAM trials. The critical points densely distribute around the mean of 189.28. **c**, Plot of the success rate, time reduction and true positive rate on the whole dataset achieved in each PAM trial. Together with (b), it shows that PAM performs stably and consistently produces high success rate.

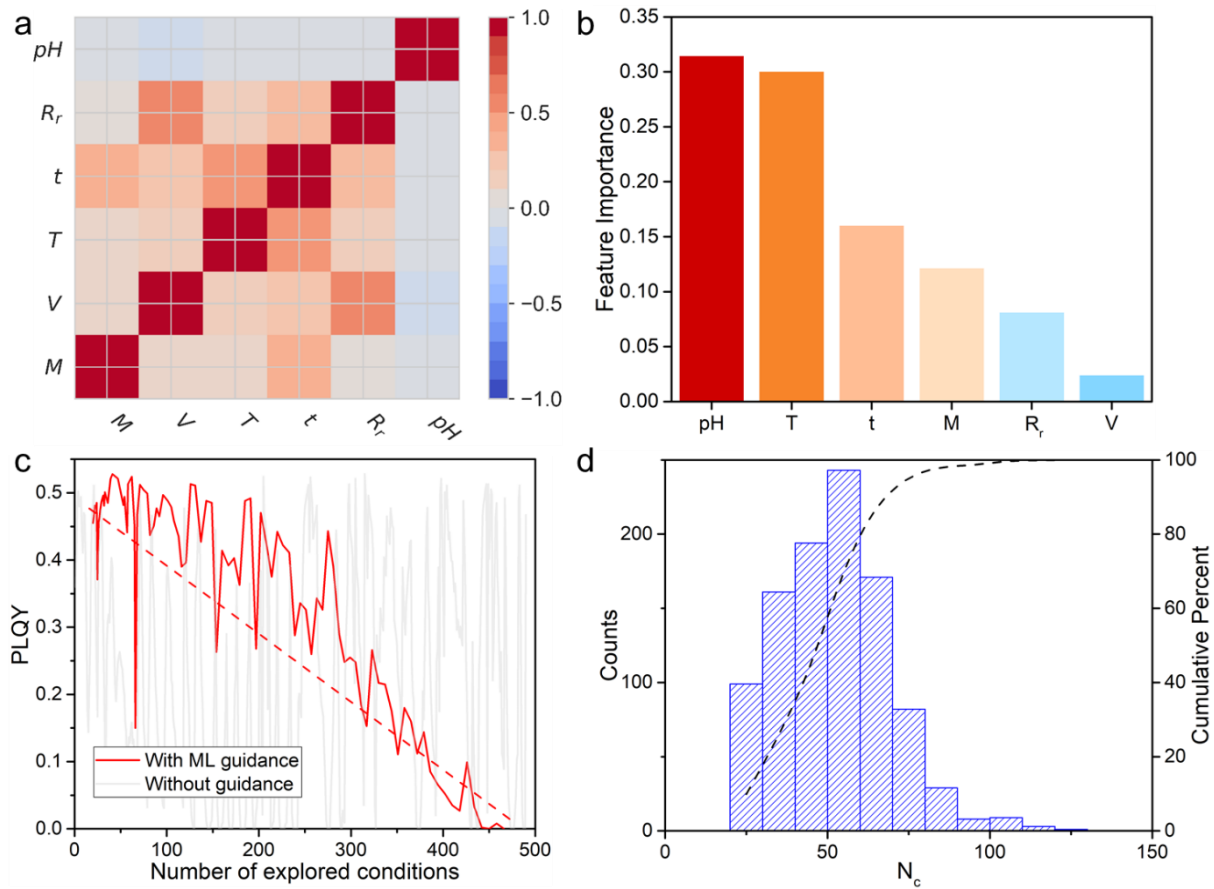

**Figure 4 | Optimization and acceleration of hydrothermal-synthesized carbon quantum dots (CQDs) with XGBoost-r and PAM. a**, The heat map of the Pearson correlation coefficient matrix among the selected features for hydrothermal-synthesized CQDs, showing generally low correlation among features. **b**, Feature importance retrieved from XGBoost-R that learns from the full dataset. The most important features are pH and T. **c**, Plot of the highest predicted/true yield versus number of explored conditions of a typical PAM trial, smoothed with mean filter of window size 3 to highlight the trend. **d**, Distribution of the critical points of 1000 PAM trials, suggesting that PAM model is 99.9% confident to find the best condition of this confined dataset within 115 experiments.

**Table 1 | Input Range of Parameters in CVD System**

| Parameter | Min | Max | Increment |
|---|---|---|---|
| Gas flow rate, $R_f$ (sccm) | 40 | 100 | 10 |
| Reaction temperature, T (°C) | 700 | 850 | 25 |
| Reaction time, t (min) | 10 | 18 | 1 |
| Distance of S outside furnace, D (cm) | 1.2 | 3.2 | 0.5 |
| Ramp time, $t_r$ (min) | 13 | 18 | 1 |
| Add NaCl | 0 or 1 | | |
| Boat configuration (Flat/Tilted) | 0 or 1 | | |

**Table 2 | Input Range of Parameters in Hydrothermal System**

| Parameter | Min | Max | Increment |
|---|---|---|---|
| pH value | 5 | 9 | 1 |
| Reaction temperature, T (°C) | 140 | 260 | 10 |
| Reaction time, t (hr) | 1 | 9 | 0.5 |
| Mass of precursor A, m (g) | 0.2 | 5 | 0.2 |
|  | 6 | 12 | 1 |
| Ramp rate, $R_r$ (°C/min) | 2, 5, 10, 15 | | |
| Solution volume, V (ml) | 10 | 60 | 5 |